\documentclass[a4paper,11pt]{article}
\usepackage{pos}
\usepackage{hyperref}
\title{Unified description of longitudinal and transverse parton dynamics in cold nuclear matter}
\author[a]{Weiyao Ke}
\affiliation[a]{Key Laboratory of Quark and Lepton Physics (MOE) \& Institute of Particle Physics, Central\\
China Normal University, Wuhan 430079, China}
\emailAdd{weiyaoke@ccnu.edu.cn}

\abstract{
We present a unified description of longitudinal momentum evolution and transverse-momentum broadening in cold nuclear matter using soft collinear effective theory with Glauber gluon interactions (SCET$_{\rm G}$). In a finite nucleus, Landau--Pomeranchuk--Migdal interference couples medium-induced energy loss to the transverse scale resolved by the observable. Our framework combines RG evolution of impact-parameter-dependent energy loss with BFKL-type rapidity evolution of the collision kernel, improved by kinematical constraints and running coupling. With a common set of cold-nuclear-matter parameters, it accounts for the main nuclear modification patterns in semi-inclusive DIS and Drell--Yan production and provides predictions for TMD hadron multiplicities at the EIC. We also discuss the one-point energy correlator in DIS, whose energy weighting suppresses the leading final-state energy-loss contribution. In the EIC kinematics studied, the residual dynamical corrections are small, leaving the nuclear modification dominated by the initial-state nuclear parton distributions.
}

\FullConference{13th International Conference on Hard and Electromagnetic Probes of High-Energy Nuclear Collisions (HardProbes2026)\\
21-26 June 2026\\
Nashville, Tennessee, USA\\}

\begin{document}
\maketitle

\section{Introduction}
\label{sec:introduction}

Partons traversing cold nuclear matter undergo both longitudinal momentum loss and transverse-momentum broadening. Semi-inclusive deep-inelastic scattering (SIDIS) and Drell--Yan (DY) production probe these medium effects in the final and initial states, respectively. In collisions with a nuclei, the finite medium path length restricts the formation time of radiation that is coherent over the medium, while a transverse-momentum measurement restricts the typical invariant mass of the probed fluctuations. These constraints couple energy loss to the transverse scale resolved by the probe. Therefore, a consistent interpretation of their nuclear modifications requires treating the two components of parton motion together.

Besides the dynamical corrections from the nuclear medium, modifications can also arise from changes in the intrinsic parton distributions of bound nucleons and nonperturbative transverse dynamics of hadron formations in the nuclear environment. We restrict this work to focus only on the dynamical corrections generated by parton-level scattering in the nucleus. 

We build on the calculation of TMD DY production in Ref.~\cite{Ke:2025TMD} and the renormalization-group (RG) analysis of collinear SIDIS in Ref.~\cite{Ke:2024RG}, and extend the phenomenology to TMD SIDIS and energy-flow observables. The main result is a common description of longitudinal attenuation and transverse broadening using the same medium parameters.

\section{Unified evolution in a finite nuclear medium}
\label{sec:framework}
For transverse momenta much smaller than the hard scale, SIDIS and DY cross sections factorize into hard, collinear, and soft functions. We work in transverse Fourier space, where $\mathbf b$ is conjugate to the transverse momentum; for SIDIS this momentum is $\mathbf p_{T,h}/z_h$. The vacuum baseline uses next-to-leading-order (NLO) matching and next-to-next-to-leading-logarithmic (NNLL) TMD evolution, with the perturbative choices and nonperturbative parametrizations of Ref.~\cite{Alrashed:2023}. Interactions with the medium are described by soft collinear effective theory with Glauber gluons, $\mathrm{SCET}_G$. The nonperturbative input consists of the path length $L$, obtained from nuclear geometry, an effective color-source density $\rho_G$ of the cold nuclear matter, and a screening mass $\xi$ of the Glauber gluon propagator.

The NLO calculation at first order in opacity contains collinear and rapidity divergences associated with medium-induced radiation. Their renormalization separates the enhanced contributions into longitudinal evolution and a rapidity-evolved transverse broadening factor. Here we summarize only the final expression for the matching coefficient after these renormalization procedures:
\begin{align}
\mathcal C_{i\rightarrow j}^{\mathrm{vac+med}}(z,b)
=e^{\rho_G L
\left[\Sigma_i(b,Y)-\Sigma_i(0,Y)\right]}\,U^{\rm vac}_{\mathrm{TMD}}
\left[C+\Delta C\right]_{i\rightarrow\ell}
\otimes U_{\ell \rightarrow k, \rm DGLAP}^{\rm vac}
\otimes M_{k\rightarrow j}^{\mathrm{med}}(z,b),
\label{eq:unified-coefficient}
\end{align}
Here $\otimes$ denotes convolution in the longitudinal momentum fraction, $C_{i\rightarrow\ell}$ and $\Delta C_{i\rightarrow\ell}$ are the finite parts of the vacuum and medium matching contributions, and $M_{k\rightarrow j}^{\mathrm{med}}$ describes the medium-induced longitudinal evolution. The broadening factor is obtained by exponentiating the product of the Glauber cross section in impact-parameter space and the areal density of color sources along the path length ($\rho_G$ L). It includes both collisional and radiative contributions.  The subtraction at $b=0$ ensures that the broadening factor preserves the transverse-momentum-integrated normalization. Equation~\eqref{eq:unified-coefficient} contains the joint evolution of longitudinal and transverse distribution, rather than treating energy loss and broadening as independent modifications.

\paragraph{Impact-parameter-dependent longitudinal evolution.} The formation time of a splitting with momentum fraction $x$ and relative transverse momentum $\mathbf k$ is $t_f\simeq 2x(1-x)E/\mathbf k^2$. Radiation with $t_f\gtrsim L$ is coherent over the entire medium. This introduces the coherence scale $\mu_L^2=2E/L$, in addition to the transverse scale $\mu_b=2e^{-\gamma_E}/b$. The logarithmically enhanced medium evolution involves the interval $\xi^2\lesssim\mu^2\lesssim \min\left\{\mu_L^2,\mu_b^2\right\}$. In the high-energy, dilute-medium limit, the coupled quark and gluon
evolution equations are given by~\cite{Ke:2024RG}:
\begin{align}
\frac{\partial M_{k\rightarrow q}}{\partial\ln\mu^2}
={}&\alpha_s\upsilon\left[
\left(4C_FC_A\frac{\partial}{\partial z}
-\frac{2C_F(2C_A+C_F)}{z}\right)M_{k\rightarrow q}
+\frac{C_FT_R}{z}M_{k\rightarrow g}\right],
\label{eq:quark-evolution}
\\
\frac{\partial M_{k\rightarrow g}}{\partial\ln\mu^2}
={}&\alpha_s\upsilon\left[
\left(4C_A^2\frac{\partial}{\partial z}
-\frac{2N_fC_F}{z}\right)M_{k\rightarrow g}
+\frac{2C_F^2}{z}\sum_{q,\bar q}M_{k\rightarrow q}\right],
\label{eq:gluon-evolution}
\end{align}
where $\upsilon\simeq\alpha_s\rho_G L^2/(4E)\ll1$, $T_R=1/2$, and the sum runs over active quark and antiquark flavors. The derivative terms generate a longitudinal shift associated with energy loss, while the off-diagonal terms couple the quark and gluon distributions. At finite $b$, the transverse phase space is restricted by $\mathbf k^2\lesssim\min\{x(1-x)Q^2,\mu_b^2\}$. If $\mu_b$ is large enough, this restriction leaves the coherent radiation region essentially unchanged. At smaller $\mu_b$, it removes part of that region and reduces the energy loss associated with that impact parameter. Therefore, the resulting longitudinal evolution depends on the transverse resolution.

\begin{figure}
\centering
\includegraphics[height=.24\textwidth]{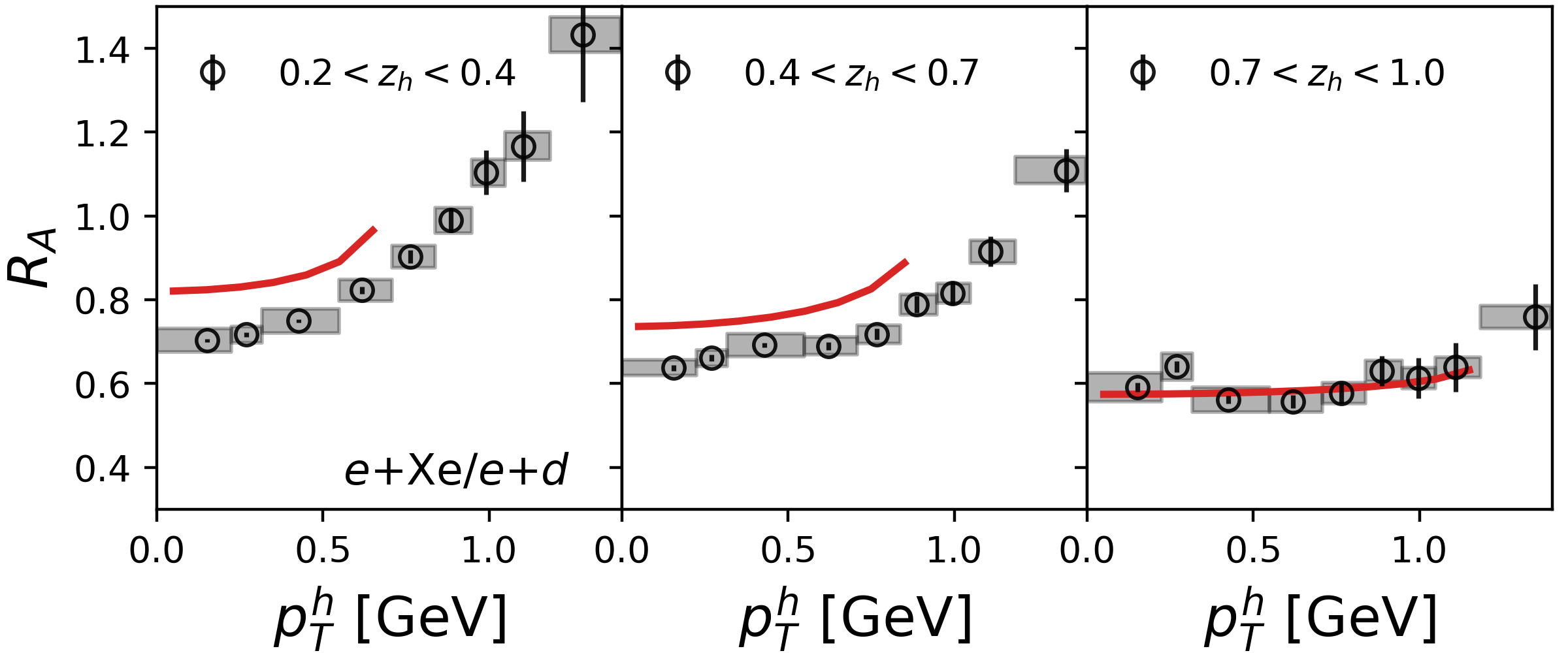}\includegraphics[height=.24\textwidth]{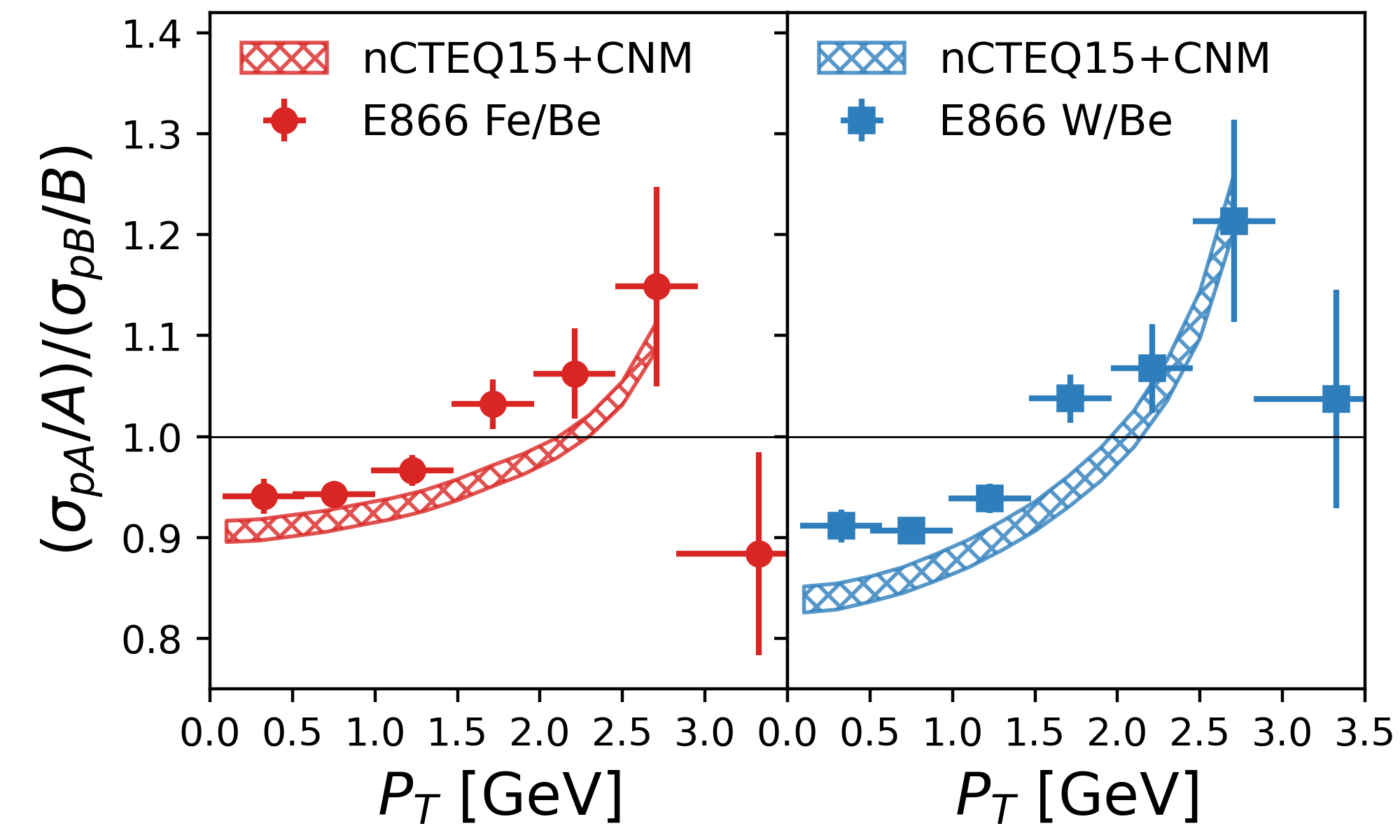}
\caption{The left panels show the HERMES $e+\mathrm{Xe}/e+d$ hadron multiplicity ratios versus $p_{T,h}$ in three $z_h$ intervals; the right panels show the E866 per-nucleon DY cross-section ratios for Fe/Be and W/Be versus dilepton transverse momentum. Both calculations use $\rho_G=0.5~\mathrm{fm}^{-3}$ and
$\xi^2=0.12~\mathrm{GeV}^2$.}
\label{fig:sidis-dy}
\end{figure}

\begin{figure}
\centering
\includegraphics[height=.32\textwidth]{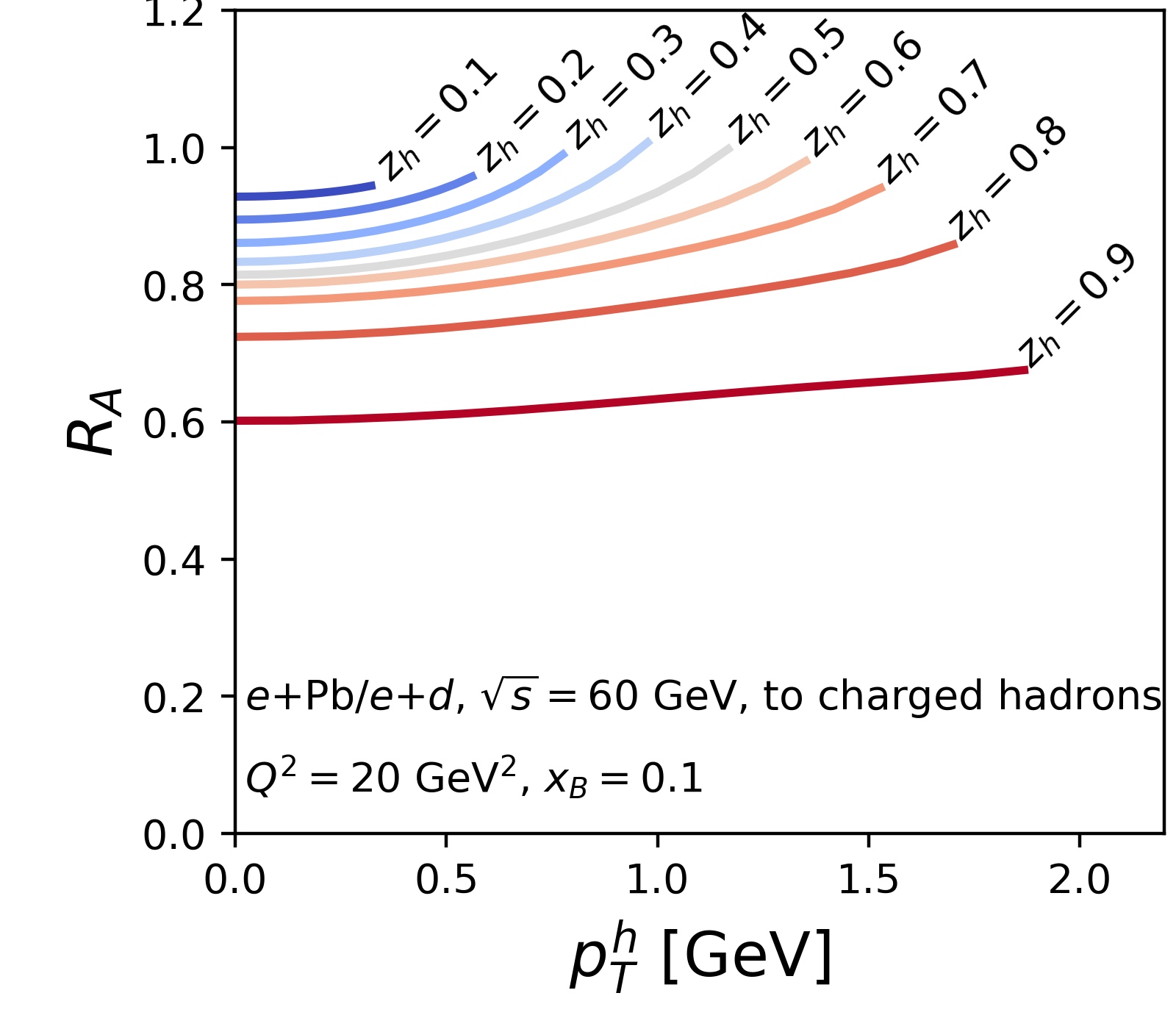}
\includegraphics[height=.32\textwidth]{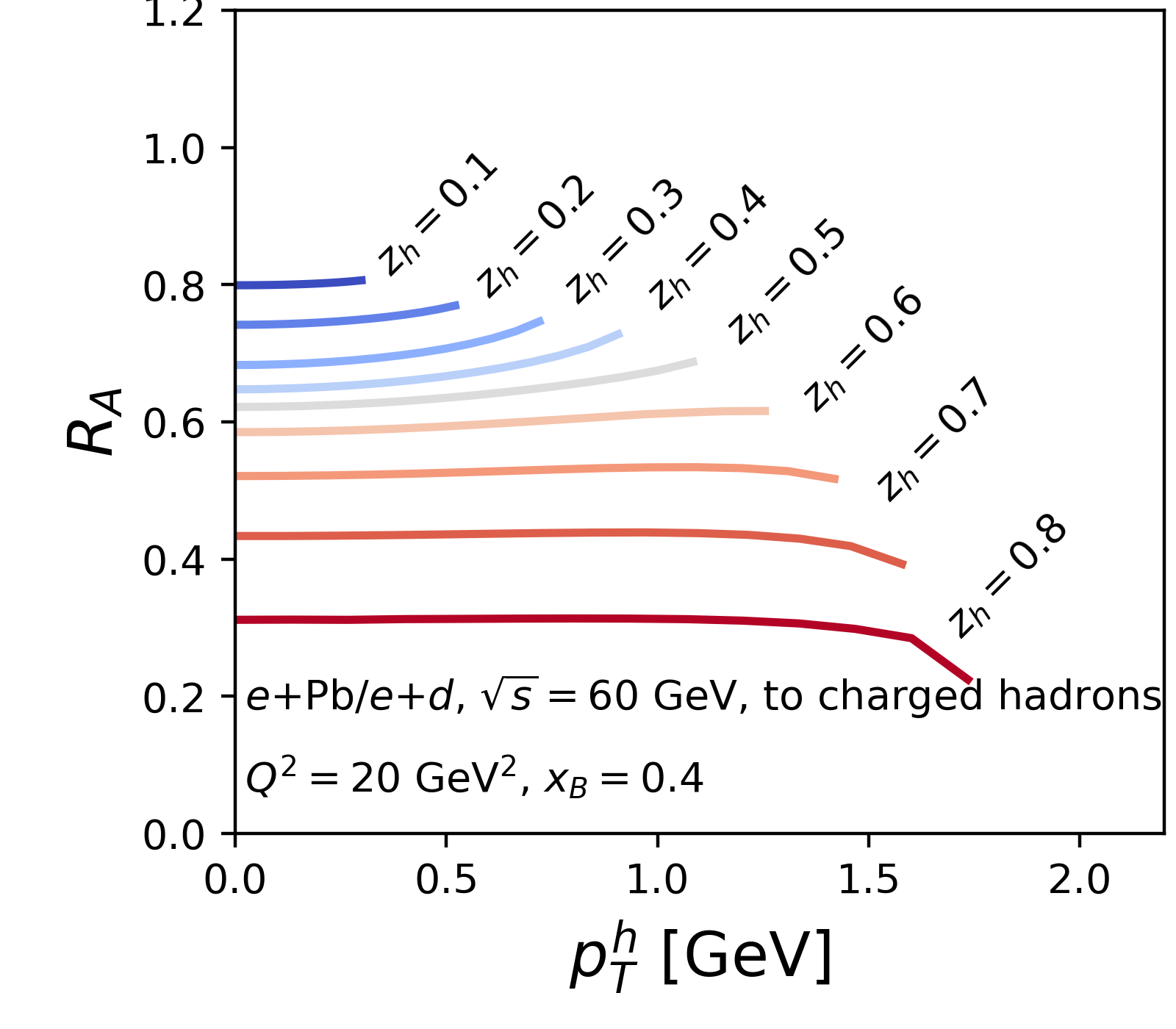}
\caption{Predicted charged-hadron multiplicity ratios for
$e+\mathrm{Pb}/e+d$ at the EIC. The collision energy is
$\sqrt{s}=60~\mathrm{GeV}$ and $Q^2=20~\mathrm{GeV}^2$;
the left and right panels correspond to $x_B=0.1$ and $0.4$,
respectively. Curves show different hadron momentum fractions
$z_h=0.1$--$0.9$.}
\label{fig:eic-sidis}
\end{figure}

\paragraph{Collisional and radiative broadening.} Defining $v(\mathbf q^2,0)=1/(\mathbf q^2+\xi^2)$, the differential cross section in $b$ space is (medium color factors and average are absorbed into the definition of $\rho_G$)
\begin{align}
\Sigma_i(b,Y)=g_s^2C_i\int\frac{d^2\mathbf q}{(2\pi)^2}
e^{-i\mathbf q\cdot\mathbf b}
v(\mathbf q^2,0)v(\mathbf q^2,Y)\,,\quad Y=\ln\left[1+\min\left\{2\xi L,
\frac{2\xi E}{\mu_b^2}\right\}\right].
\label{eq:rapidity-range}
\end{align}
Here $C_i=C_F$ or $C_A$ for a quark or gluon. $E$ is the parton energy in the rest frame of the nucleus. At $Y=0$, $\Sigma_i(b,0)$ only generates collisional broadening. A BFKL rapidity evolution of $v$ resums radiative corrections to the scattering process. The rapidity window to be resummed is bounded by both the nuclear length $(\xi L)$ and the transverse-momentum  measurement  $(2\xi E/\mu_b^2)$. However, the unconstrained leading-order BFKL evolution produces excessively strong radiative broadening. We improve the kernel by implementing kinematical constraints following Ref.~\cite{Deak:2019}:
\begin{align}
\hskip-.5em\frac{\partial v(\mathbf k^2,Y)}{\partial Y}
=\frac{C_A}{\pi}\int\frac{d^2\mathbf q}{\pi\mathbf q^2}
\Bigg[
\alpha_{s,1}\Theta\left(\frac{\mathbf k^2}{e^Y\mathbf q^2}-1\right)v\left(|\mathbf k+\mathbf q|^2,
Y-\Delta Y\right)-\alpha_{s,2}\Theta\left(\frac{\mathbf k^2}{\mathbf q^2}-1\right)
v(\mathbf k^2,Y)\Bigg],
\label{eq:constrained-bfkl}
\end{align}
with $\Delta Y=\ln\max\left\{1,\frac{\mathbf q^2}{\mathbf k^2}\right\}$. Together with running coupling prescription motivated by Ref.~\cite{Kovchegov:2007}, $\alpha_{s,1}={}\alpha_s(\mathbf q^2+\xi^2)$ and $\alpha_{s,2}=\alpha_s(\mathbf q^2+\xi^2)\alpha_s(|\mathbf k+\mathbf q|^2+\xi^2) / \alpha_s(\mathbf k^2+\xi^2)$,  the growth of the radiative contribution is moderated.

\section{SIDIS and Drell--Yan phenomenology}
\label{sec:phenomenology}
For SIDIS, we study the nuclear modification factor $R_A(x_B,Q^2,z_h,p_{T,h})$ between lead and deuteron, defined as the ratio of hadron multiplicities per DIS event. We include collinear nuclear parton distributions in the calculations, while using the vacuum nonperturbative transverse input without introducing a separate fitted nuclear TMD parametrization.  For Drell--Yan, we compute the mass-number-normalized differential cross sections for different nuclei. Figure~\ref{fig:sidis-dy} compares the calculations with the HERMES SIDIS data~\cite{HERMES:2007} and the E866 DY data~\cite{E866:1999}. Both processes use $\rho_G=0.5~\mathrm{fm}^{-3}$ and $\xi^2=0.12~\mathrm{GeV}^2$. In SIDIS, longitudinal evolution suppresses the yield at large $z_h$, while transverse broadening redistributes the yield toward larger $p_{T,h}$. The interplay of these effects produces a rising multiplicity ratio with $p_{T,h}$ at smaller $z_h$ and stronger attenuation at larger $z_h$. The DY ratios also rise with transverse momentum, reflecting initial-state broadening. The common parameter set accounts for the main trends in both processes, providing a consistency test of the dynamical cold nuclear matter (CNM) effects calculation.

Using the same parameters, Fig.~\ref{fig:eic-sidis} presents predictions for charged-hadron production in $e+\mathrm{Pb}$ relative to $e+d$ at $\sqrt{s}=60~\mathrm{GeV}$ and $Q^2=20~\mathrm{GeV}^2$. The nuclear attenuation is stronger at $x_B=0.4$ than at $x_B=0.1$, and increases with $z_h$. This dependence follows from the outgoing parton energy in the nuclear rest frame, $E\simeq Q^2/(2m_Nx_B)$: increasing $x_B$ at fixed $Q^2$ lowers the energy and enhances the relative energy-loss effect. The $p_{T,h}$ dependence provides an additional constraint on broadening and on the transverse-scale dependence of the longitudinal evolution.

\paragraph{Energy-weighted observables} The one-point energy correlator (OPEC) measures energy flow relative to the incoming proton direction in the Breit frame, with angular variable $\tau=(1+\cos\theta)/2$. Its back-to-back limit, $\tau\ll1$, admits TMD factorization~\cite{Li:2021EEC}; its nuclear dependence has also been studied in Ref.~\cite{Fu:2026OPEC}. Here we estimate the dynamical CNM correction using the parameters constrained by SIDIS and DY. Energy weighting removes the leading dependence on collinear fragmentation through the momentum sum rule $\sum_h\int_0^1 dz_h\,z_h d_{h/j}(z_h,\mu)=1$. For a quark-initiated channel, the corresponding energy-weighted jet function reduces to a partonic moment,
\begin{align}
J_q(b)
=\int_0^1 dx\,x\left[
\mathcal I_{q/q}\left(x,\frac{b}{x},\mu,\nu\right)
+\mathcal I_{g/q}\left(x,\frac{b}{x},\mu,\nu\right)
\right],
\label{eq:energy-weighted-jet}
\end{align}
where $\mathcal I_{j/q}$ are the collinear matching kernels and
$\nu$ is the rapidity renormalization scale. The energy-weighted sum over the quark and gluon channels cancels the collinear divergences in both the vacuum and medium contributions. The leading energy-loss enhancement is consequently suppressed, although finite transverse-broadening corrections remain.

\begin{figure}
\centering
\includegraphics[height=.28\textwidth]{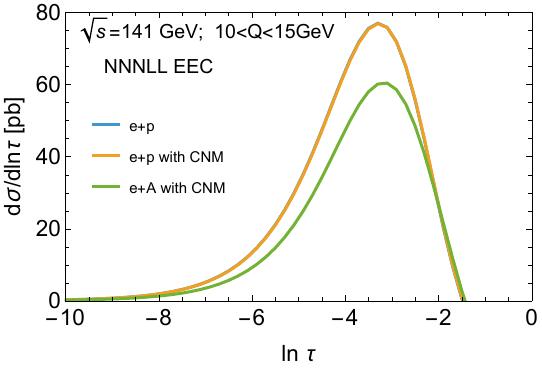}
\includegraphics[height=.28\textwidth]{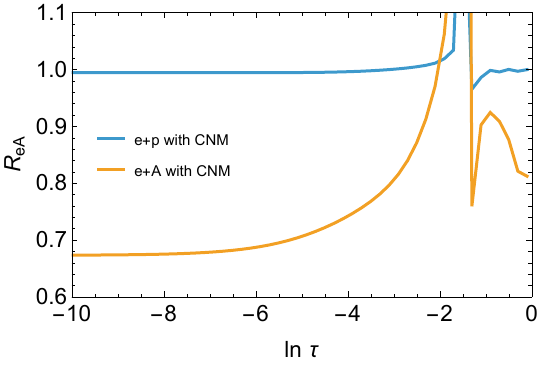}
\caption{The left panel compares $d\sigma/d\ln\tau$ for the $ep$ baseline, a CNM-only calculation using proton PDFs, and the full $eA$ calculation including nPDFs. The right panel shows the corresponding ratios to the $ep$ baseline. The collision energy is $\sqrt{s}=141~\mathrm{GeV}$ and $10<Q<15~\mathrm{GeV}$.}
\label{fig:opec}
\end{figure}

Figure~\ref{fig:opec} shows the prediction at
$\sqrt{s}=141~\mathrm{GeV}$ and $10<Q<15~\mathrm{GeV}$. In the main back-to-back region shown, the residual dynamical correction is below the percent level, while the nuclear modification reaches approximately $30\%$ and is dominated by the initial-state nPDF input. Together with identified-hadron SIDIS, the OPEC offers a way to distinguish nuclear parton distributions from final-state parton evolution.

\section{Summary}
\label{sec:summary}

We have presented a unified description of longitudinal and transverse parton evolution in cold nuclear matter. Finite-length interference and transverse-momentum constraints generate an
impact-parameter-dependent longitudinal evolution, while
rapidity evolution of the Glauber collision kernel accounts for
radiative broadening. Kinematical constraints and running coupling moderate the BFKL evolution and allow the main SIDIS and DY modification patterns to be described with the same CNM parameters. The resulting predictions motivate combined collinear and TMD measurements to test our understanding of medium correction to parton evolution. Finally, energy-weighted observables such as the OPEC provide complementary information by suppressing the leading final-state energy-loss contribution.

{\bf Acknowledgments:} W.K. is supported by the National Natural Science Foundation of China under Grant No.\ 12575140.


\begin{thebibliography}{99}

\bibitem{Ke:2025TMD}
W.~Ke, J.~Terry and I.~Vitev,
\textit{Toward a first-principles description of transverse momentum
dependent Drell--Yan production in proton-nucleus collisions},
JHEP \textbf{02} (2025) 102,
arXiv:2408.10310 [hep-ph].

\bibitem{Ke:2024RG}
W.~Ke and I.~Vitev,
\textit{Understanding parton evolution in matter from
renormalization group analysis},
Phys.\ Lett.\ B \textbf{854} (2024) 138751,
arXiv:2301.11940 [hep-ph].

\bibitem{Alrashed:2023}
M.~Alrashed, Z.-B.~Kang, J.~Terry, H.~Xing and C.~Zhang,
\textit{Nuclear modified transverse momentum dependent parton
distribution and fragmentation functions},
arXiv:2312.09226 [hep-ph].

\bibitem{Deak:2019}
M.~Deak, K.~Kutak, W.~Li and A.~M.~Stasto,
\textit{On the different forms of the kinematical constraint in BFKL},
Eur.\ Phys.\ J.\ C \textbf{79} (2019) 647.

\bibitem{Kovchegov:2007}
Y.~V.~Kovchegov and H.~Weigert,
\textit{Quark loop contribution to BFKL evolution:
Running coupling and leading-$N_f$ NLO intercept},
Nucl.\ Phys.\ A \textbf{789} (2007) 260--284.

\bibitem{HERMES:2007}
A.~Airapetian \textit{et al.} [HERMES],
\textit{Hadronization in semi-inclusive deep-inelastic scattering
on nuclei},
Nucl.\ Phys.\ B \textbf{780} (2007) 1--27.

\bibitem{E866:1999}
M.~A.~Vasiliev \textit{et al.} [FNAL E866/NuSea],
\textit{Parton energy loss limits and shadowing in
Drell--Yan dimuon production},
Phys.\ Rev.\ Lett.\ \textbf{83} (1999) 2304.

\bibitem{Li:2021EEC}
H.~T.~Li, Y.~Makris and I.~Vitev,
\textit{Energy-energy correlators in deep inelastic scattering},
Phys.\ Rev.\ D \textbf{103} (2021) 094005,
arXiv:2102.05669 [hep-ph].

\bibitem{Fu:2026OPEC}
Y.~Fu, Z.-B.~Kang, J.~Penttala and Y.~Zhou,
\textit{Exploring nuclear modification using one-point energy
correlator at the electron-ion collider},
JHEP \textbf{06} (2026) 247,
arXiv:2512.16847 [hep-ph].

\end{thebibliography}
\end{document}